\documentclass[conference]{IEEEtran}

\usepackage[colorinlistoftodos,prependcaption,textsize=footnotesize, color=red]{todonotes} 

\usepackage{import}
\usepackage{graphicx}
\usepackage{xcolor}
\usepackage{siunitx}
\usepackage{multirow}
\usepackage{booktabs}
\usepackage{balance}
\usepackage{tikz}
\usepackage{import}
\usepackage{paralist}
\usepackage{subcaption}

\usepackage{xspace}

\newcommand{\eg}{e.\,g.,\xspace}

\newcommand{\ie}{i.\,e.,\xspace}

\usepackage{makeidx}

\newcommand{\txkr}[1]{\tikz[baseline=-.75ex, anchor=center]{\draw circle (1ex); \node[outer sep=0, inner sep=0] {#1};}}

\begin{document}

\title{An SDN-based Approach For Defending Against Reflective DDoS Attacks}

\author{
    \IEEEauthorblockN{Thomas Lukaseder, Kevin St\"olzle, Stephan Kleber, Benjamin Erb, Frank Kargl}
	\IEEEauthorblockA{Institute of Distributed Systems\\
	Ulm University, Germany\\
	\{firstname\}.\{lastname\}@uni-ulm.de}
}

\maketitle

\begin{abstract}

Distributed Reflective Denial of Service (DRDoS) attacks are an immanent threat to Internet services. The potential scale of such attacks became apparent in March 2018 when a memcached-based attack peaked at 1.7 Tbps.
Novel services built upon UDP increase the need for automated mitigation mechanisms that react to attacks without prior knowledge of the actual application protocols used.
%
With the flexibility that software-defined networks offer, we developed a new approach for defending against DRDoS attacks; it not only protects against arbitrary DRDoS attacks but is also transparent for the attack target and can be used without assistance of the target host operator. 
The approach provides a robust mitigation system which is protocol-agnostic and effective in the defense against DRDoS attacks.

\end{abstract}



\section{Introduction}


Distributed Denial of Service (DDoS) attacks are one of the key threats to network security.
A new record for the extent of such an attack was set in March 2018 when an attack exceeded 1.7 Tbps of traffic~\cite{memcached}.
This ``memcached attack'' is a Distributed Reflective Denial of Service (DRDoS) that does not attack the target directly but instead send request packets to an exploitable third party service (\ie the reflector) with a spoofed sender IP address. 
The responses of the third party server are then sent to the actual attack target and cause overloading.
Protocols with response messages that are significantly larger than request messages are particularly well suited for these attacks due to amplification effects. The nature of these attacks requires services that work without an established connection between client and server. 
In a recent analysis~\cite{DoSThreat} in 2017, Jonker et al. found that 99.27\% of all DRDoS attacks are using the protocols NTP, DNS, CharGen, SSDP, and RIPv1; all of these are based on UDP and possess a registered or well-known port number just like the newer memcached attack.
Messages received in a DRDoS attack are hard to differentiate from benign traffic, as they conform to the protocol specification. The reflectors are correctly handling requests they deem to be legitimate. 
However, the lack of a request for the responses is a characteristic of reflective attacks that cannot be masked. Due to the stateless design of UDP, additional mechanisms on the application layer are often used to attribute a response to a corresponding request. Tracking such request/response mappings within a network is infeasible for several reasons including scalability issues and the necessity to probe application layer messages.

To this end, we introduce a lightweight NAT-based packet filtering approach that reliably mitigates most common reflective attacks without disrupting regular connections.
This approach does not require participation of the target host administrator and can therefore be offered by ISPs \emph{as-a-service} to any of their customers.
Software-defined networking (SDN) provides the flexibility to build and deploy such a system without interfering with the existing network infrastructure since the system can be added to the SDN controller. 

The remainder of this paper is organized as follows:
Section~\ref{sec:related} outlines previous mechanisms for DRDoS mitigation.
In Section~\ref{sec:arch}, we introduce the architecture of our detection and mitigation system. Section~\ref{sec:impl} explores the implementation details. Section~\ref{sec:conclusion} concludes the paper and outlines our plans for future development.


\section{Related Work}
\label{sec:related}

The inherent characteristics of DRDoS attacks lead to unique challenges for their mitigation.
To provide an effective defense for a potential target of such attacks, the mitigation mechanism must be deployed in the network of the target host.
We discuss previous defense approaches against DDoS and DRDoS attacks, which proposed solutions to the specific challenges of this kind of attack.

A review of related work showed that defense against DRDoS attacks can be divided into multiple tasks: 
(1a) monitoring and data collection of network traffic, (1b) interpretation of this data and detection of an ongoing attack, and 
(2) prevention or mitigation of the attack.
Depending on their primary tasks, existing solutions vary widely in their assumptions and requirements.
As we take into account high bandwidth networks, we also assess the level of scalability of existing solutions on the task level.
Especially the data collection has high impact on scalability, since each packet in the network traffic must be regarded and processed in a timely manner.
This task runs continuously for all traffic during the attack-free operation of the network.
Therefore, the interpretation and attack detection task becomes critical for the latency within the network.
Passive monitoring does not suffer from this limitation as traffic is not affected during detection.
The final task for an ongoing attack once detected is mitigating the impact of the attack without interfering with legitimate traffic.

For example \emph{PacketScore}, by Kim et al.~\cite{PacketScoreNew}, and its extension
\emph{ALPi}, proposed by Ayres et al.~\cite{ALPi}.
Both use packet statistics to compare each packet to benign and attack traffic and rate each packet. Packets that are similar to attack traffic get discarded.
The approach requires active network monitoring and, therefore, the data collection impacts the latency of all traffic.
The resource requirements are considerably higher for PacketScore than for other related work, in particular for high bandwidth networks.
One of ALPi's latency-optimized detection mechanisms, the lightweight leaky bucket, yields 2.82\,\% false negatives at a false positive ratio of less than 0.1\,\%.
The approach was designed to defend against general network attacks and should be applicable to DRDoS attacks with a modified set of packet attributes.


The approach of Wei et al.~\cite{wei2013rank} uses traffic statistics.
It correlates the packet rates for all flow pairs passing the same router.
Legitimate traffic is assumed not to exhibit any such correlation, while flows belonging to an attack linearly correlate in their packet rate.
However, the calculation of pairwise correlation coefficients for each pair of flow makes the detection very costly.
According to Wei et al., the quality of the DRDoS detection has a false negative ratio of 0.18,\% and a false positive ratio of 0.10\,\%.
Gao et al.~\cite{gao2016machine} conducted their own evaluation of Wei et al.~\cite{wei2013rank}, which showed a detection rate of 96\,\% but a false positive ratio as high as 30\,\%.

In our earlier work~\cite{scoringBasedIpFiltering}, we employ traffic statistics to detect attacks and then mitigate them using SDN.
This method probes the service of the target host to estimate its load. 
If the service becomes unavailable, the client hosts that produce most of the target's load are scored as suspicious and their traffic is redirected.
Since passive monitoring is employed, no latency is introduced by the data collection.
The detection via scoring is very lightweight; therefore it scales well to high-speed networks. However, a basic analysis of the whole traffic data is still needed.
Redirecting suspicious traffic to a CAPTCHA server prevents to cut-off false positives from the service.
Depending on the scenario, we determined between 5\,\% and 18\,\% false positives in the identification of attackers.
The approach originally was not designed for reflective but direct distributed-denial-of-service attacks, but can be applied to both with only minimal changes.

Other approaches exist that have stronger assumptions about the network environment than the presented approaches and our work.
For example, L-RAD~\cite{RAD} proposes an active message authentication by the deep integration of the target host into the detection mechanism.
To incorporate multiple defenses in a flexible and scalable way, Mahimkar et al.~\cite{dFence} propose the architecture \emph{dFence} to deploy arbitrary detection and mitigation approaches.
The goals of these works are distinctively different from our proposal, therefore we do not discuss them further.

A common mitigation mechanism for reflective DDoS attacks is to completely block the port of the service that gets abused. 
This successfully mitigates the attack; however, any legitimate traffic using the same port (e.\,g., legitimate DNS requests during a DNS-based reflective attack) is also affected.

The mitigation of attacks in all these models depends on the analysis of all packets going through the network node where the mitigation is located, which is very costly. Therefore, we developed a mitigation mechanism that does not need to analyze any attack packets during mitigation except for the usual layer-3 forwarding rules.


\section{Architecture}
\label{sec:arch}



\begin{figure*}
    \small
	\centering
    \def\svgwidth{1.3\columnwidth}
    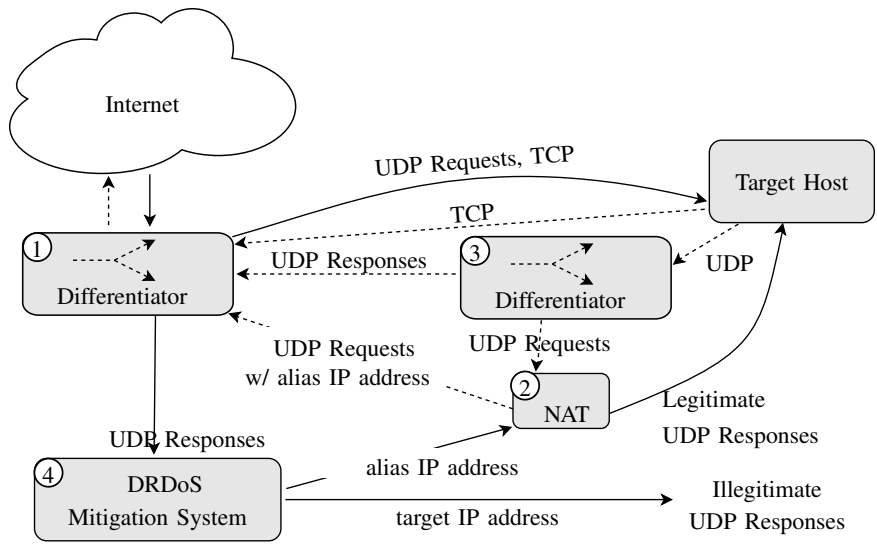
	\caption{Sketch of the mitigation mechanism.}
	\label{fig:differentiationApproachStep2}
\end{figure*}

The building blocks of our mitigation architecture are illustrated in Figure~\ref{fig:differentiationApproachStep2} and referenced throughout the text in the encircled numbers \txkr{1} to \txkr{4}.
The mitigation is based on NAT packet filtering, which is implemented in SDN.
If a detection system reports a DRDoS attack, our mitigation is activated automatically.

\txkr{1} In general, there are four types of messages receivable by the target host: legitimate requests and responses and illegitimate (malicious) requests and responses. During a DRDoS attack, illegitimate responses must be filtered while during other DDoS attacks malicious requests must be filtered. Therefore, a separation of requests and responses is the first step to mitigate a reflective attack. DRDoS attacks are typically performed using UDP, therefore only UDP packets are analyzed by the DRDoS mitigation system. As a result, all UDP responses are forwarded to the DRDoS mitigation system, while all UDP requests and TCP packets are forwarded to the destination or alternatively to a DDoS mitigation system.

\txkr{2} The DRDoS mitigation system needs to classify incoming responses either as legitimate or illegitimate. The basic assumption used to differentiate legitimate responses from illegitimate responses is that a response can only be legitimate if the target host sent a corresponding request beforehand. 

A very simple solution to determine whether a corresponding request has been sent is the use of Network Address Translation (NAT)~\cite{RFC3022, RFC4787}. We use an adapted form of NAT.
During an attack, the NAT is activated and the source IP address of the attack target (Fig.~\ref{fig:differentiationApproachStep2}: target IP address) is replaced by an alias IP address in \emph{outgoing UDP-based requests}.
Therefore, all responses created upon these requests are sent to the alias IP address. The destination IP address of incoming responses destined to the alias IP address is replaced by the corresponding target IP address and forwarded to the target host. This mechanism can be used to effectively distinguish legitimate responses from illegitimate ones without observing the attack traffic. Only the benign UDP \emph{requests} from the attack target have to be tracked by the NAT.
Legitimate responses contain the alias IP address as their destination IP address, while illegitimate responses contain the target IP address as their destination IP address. TCP packets and incoming UDP requests are forwarded to the target as usual. The NAT-based solution scales very well for very high attack packet rates as the attack traffic does not have to be observed.

\txkr{3} A second differentiator separates requests from responses to be able to use NAT only for outgoing requests, but not for outgoing responses. Using this concept, the target host does not have to be aware of the DRDoS mitigation mechanism. All required tasks to mitigate the attack are implemented by the network address translator and the differentiator. \txkr{4} The DRDoS mitigation is a simple forwarding unit that discards all UDP packets addressed to the target IP address. It is imperative that the alias IP address is not easily guessable, so it is not easily possible for attackers to switch their attack to the alias IP address. Since the alias IP address is only visible to UDP-based services utilized by the target host,
(and these services cannot know that this request comes from the target as the alias IP address should not be known)
the alias IP address can only be discovered by the attacker if the attacker can observe the network traffic at the target. For additional security, if an attacker might still be able to guess the alias IP address, the address could also be changed regularly (with a grace period where both IP addresses are accepted to allow for open connections to finish) and thereby implementing a moving target defense.

The effectiveness of this approach depends on the distinctiveness of responses and requests. One possibility to differentiate responses from requests would be to analyze the payload, possibly containing a response flag, or similar static substrings. Using payload analysis does not scale well and requires protocol-specific knowledge. However, most services that are used for reflective attacks have a default port (\eg 53 for DNS) that can be used instead~\cite{DoSThreat}. 
Our detection mechanism is able to extract and identify the application protocol port from the UDP header and to forward it to the mitigation system.
This approach is application protocol-agnostic and supports any UDP-based request/response protocol that follows a client/server design.

\section{Implementation}
\label{sec:impl}

\begin{table}
	\centering
	\caption[Flow entries of the combined implementation.]{Flow entries of the combined solution of the differentiator-based approach. $\mathit{IP}_t$: IP address of the target host. $\mathit{IP}_a$: alias IP address used by the NAT component. $p_a$: default port of the attack.}
	\begin{tabular}{|c|c|c|c|c|l|}
		\hline
		\multicolumn{5}{|c|}{Match Fields} & \multirow{4}{*}{\textbf{Action}}\\
		\cline{1-5}
		\multirow{2}{*}{\textbf{EtherType}} & \multicolumn{2}{|c|}{IPv4} & \multicolumn{2}{|c|}{UDP} & \\
		\cline{2-5}
		& \textbf{SRC} & \textbf{DST} & \textbf{SRC} & \textbf{DST} & \\
		\hline
		0x0800 (IPv4) & $\mathit{IP}_t$ & $*$ & $*$ & $p_a$  & $\Rightarrow$ CONTROLLER \\
		0x0800 & $\mathit{IP}_t$ & $*$ & $p_a$ & $*$  & $\Rightarrow$ TARGET \\
		0x0800 & $*$ & $\mathit{IP}_t$ & $*$ & $p_a$  & $\Rightarrow$ TARGET \\
		0x0800 & $*$ & $\mathit{IP}_t$ & $p_a$ & $*$  & DROP \\
		0x0800 & $*$ & $\mathit{IP}_a$ & $p_a$ & $*$  & $\Rightarrow$ CONTROLLER \\
		\hline
		0x0806 (ARP) & $*$ & $*$ & $*$ & $*$  & $\Rightarrow$ CONTROLLER \\
		\hline
	\end{tabular} 
	\label{tab:sdn-downwards}
\end{table}

\begin{table*}
	\centering
	\caption[Flow entries of the SDN switch-only implementation.]{Flow entries of the SDN switch-only implementation of the differentiator-based approach. $\mathit{IP}_t$: IP address of the target host. $b(\mathit{IP}_v)$: Broadcast address of the target host's subnetwork. $\mathit{IP}_a$: alias IP address used by the NAT component. $\mathit{MAC}_v$: MAC address of the target host. $p_a$: default port of the attack.}
	\begin{tabular}{|c|c|c|c|c|c|c|l|}
		\hline
		\multicolumn{7}{|c|}{Match Fields} & \multirow{4}{*}{\textbf{Action}} \\
		\cline{1-7}
		\multirow{2}{*}{\textbf{EtherType}} & \multicolumn{2}{c|}{ARP} & \multicolumn{2}{c|}{IPv4} & \multicolumn{2}{c|}{UDP} & \\
		\cline{2-7}
		 & \textbf{OP} & \textbf{TPA} & \textbf{SRC} & \textbf{DST} & \textbf{SRC} & \textbf{DST} & \\
		\hline
		0x0800 (IPv4) & $*$ & $*$ & $\mathit{IP}_t$ & $*$ & $*$ & $p_a$  & \begin{tabular}{@{}l@{}}
																	set-field IPV4\_SRC = $\mathit{IP}_a$ \\
																	 $\Rightarrow$ TARGET 
																\end{tabular} \\
		\hline
		0x0800 & $*$ & $*$  & $\mathit{IP}_t$ & $*$ & $p_a$ & $*$  &  $\Rightarrow$ TARGET  \\
		0x0800 & $*$ & $*$  & $*$ & $\mathit{IP}_t$ & $*$ & $p_a$  &  $\Rightarrow$ TARGET  \\
		0x0800 & $*$ & $*$  & $*$ & $\mathit{IP}_t$ & $p_a$ & $*$  & DROP \\
		\hline
		0x0800 & $*$ & $*$  & $*$ & $\mathit{IP}_a$ & $p_a$ & $*$  & \begin{tabular}{@{}l@{}}
																	set-field IPV4\_DST = $\mathit{IP}_t$ \\
																	 $\Rightarrow$ TARGET 
																\end{tabular} \\
		\hline
		0x0806 (ARP) & 1 & $\mathit{IP}_a$  & $*$ & $*$ & $*$ & $*$  & \begin{tabular}{@{}l@{}}
																	set-field ETH\_SRC = $\mathit{MAC}_v$ \\
																	set-field ETH\_DST = ff:ff:ff:ff:ff:ff \\
																	set-field ARP\_OP = 2 \\
																	set-field ARP\_SPA = $\mathit{IP}_a$ \\
																	set-field ARP\_SHA = $\mathit{MAC}_v$ \\
																	set-field ARP\_TPA = $b(\mathit{IP}_v)$ \\
																	set-field ARP\_THA = ff:ff:ff:ff:ff:ff \\
																	 $\Rightarrow$ TARGET 
																  \end{tabular} \\
		\hline
	\end{tabular} 
	\label{tab:sdn-switchonly}
\end{table*}

The design goal of our system is the ease of deployment in a given software-defined network with low overhead and without changing the network topology. We base our system on our prior work introduced in 2017 \cite{scoringBasedIpFiltering} and extended in 2018 \cite{slowdos}.
Two components are required to implement this system: for one, a monitoring system that can handle the traffic load and that meets our feature requirements, and second, an SDN controller that works with our hardware, allows for rapid prototyping, and works well with our network monitor. 
Based on these requirements, we chose the Bro Network Monitor\footnote{https://bro.org} to detect attacks based on prior research and Ryu\footnote{https://github.com/osrg/ryu} as SDN controller.
Moreover, these tools are already interoperable by the well-established Bro Client Communications Library ``Broccoli''.

The Ryu application of the prototype provides the implementation of the mitigation mechanism.
Bro informs Ryu of an attack supplemented by the target host's IP address and the default port $p_a$ of the attack. Blocking or redirecting traffic of single clients is realized by specifying a flow entry for every reported IP address of targets under attack. We implemented the NAT component in two different ways. One solution utilizes the SDN switch to forward packets based on their header fields and then employs the Ryu controller to actively modify packets before they are forwarded. The SDN rules for this implementation can be found in Table~\ref{tab:sdn-downwards}.
This implementation relies on mandatory OpenFlow 1.3 features and, therefore, works with any switch that implements OpenFlow 1.3 correctly.
Our second implementation makes use of optional OpenFlow 1.3 features.
The utilized rules are shown in Table~\ref{tab:sdn-switchonly}.
This variant works without the involvement of the SDN controller in modifying packets and is therefore faster and more scalable.
However, not all switches implementing OpenFlow 1.3 have the capability to run this implementation.

Our hardware features an HPE FlexFabric 5920 switch which supports OpenFlow 1.3. The web server (attack target) and the monitoring system (Ryu and Bro) run on separate systems each with the following specifications: CPU with 4x3.10Ghz\footnote{Intel® Xeon® Processor E3-1220 v3} and 17GiB of memory. The attacks are run from three replay servers with a 6x2.40 GHz CPU\footnote{Intel® Xeon® Processor E5-2630 v3} and 125GiB of memory each. With this hardware, a throughput of 6.5 Gbps could be achieved and attacks at that rate were successfully mitigated.


\section{Conclusion}
\label{sec:conclusion}

DRDoS attacks cause vast economic damage. It is imperative to detect and mitigate these attacks early before the attack target is impaired. We designed and implemented a new mitigation mechanism that can reliably detect and mitigate arbitrary DRDoS attacks as long as the underlying protocol uses UDP and a fixed server port. The mitigation system described in this paper makes use of basic SDN features and can be implemented in any network with switches that are OpenFlow 1.3 capable. The mitigation is entirely network-based (\ie independent of the attack target) and is fully transparent to the target. The architecture description in this paper should allow for the implementation of this defense system in any SDN-based network.

In this paper, we only focused on the defense side of reflective attacks. For future work, we will work on improvements to the detection mechanims.



\section*{Acknowledgment}

This work was supported in the bwNET100G+ project
by the Ministry of Science, Research and the Arts Baden-
W\"urttemberg (MWK). The authors alone are responsible for
the content of this paper.

\bibliographystyle{IEEEtranS}
\bibliography{literature}
\end{document}